# The Temporal Gearbox

B. Jalali*, J. C.K. Chan, A. Mahjoubfar, D. Solli, M. H. Asghari

University of California, Los Angeles, 420 Westwood Plaza, Los Angeles, 90095, USA

## ABSTRACT

We are inspired by mechanical gearboxes and demonstrate its analog counterpart in optics. We use nonlinear dispersion modes as "gears" to overcome the mismatch between the ultrahigh speed of optical data and the much slower sampling rate of electronic digitizers and processors. We delineate the mathematical foundations and show its utility in ultrafast optical measurements and digital image compression.

## 1. INTRODUCTION

In optical communication and sensing systems, the signal is converted by a photodetector to an analog electrical signal and then digitized with an analog-to-digital converter (ADC). The fundamental and practical limit to performance of all such systems is the speed and the accuracy of the ADC. Specifically, there exists a sampling mismatch between the speeds of signal and that of the ADC. To alleviate this problem, Photonic Time Stretch, an analog slow motion technique, was introduced in the 1990s[1,2]. In photonic time stretch, an optical fiber with large chromatic dispersion is used to slow down wideband optical signals that are encoded on the optical spectrum. The immediate yet powerful consequence of the analog gearbox in bridging the speed mismatch between fast optical signals and slow ADCs is the ability to perform ultrafast optical measurements in a single-shot fashion.

The successes of the analog gearbox has been very well-documented, ranging from the discovery of optical rogue waves[3] and the phenomenon of soliton explosions in fibers[4], to the first real-time investigation of laser mode-locking dynamics[5] and relativistic electron bunching in synchrotron radiation[6]. It also enabled single-shot performance in stimulated Raman spectroscopy (SRS)[7] as well as a milestone demonstration of single-shot SRS measurements of carbon-hydrogen bond in chemicals at MHz frame rates[8]. When combined with deep learning, label-free detection of cancer cells in blood was also achieved with record accuracy[9].

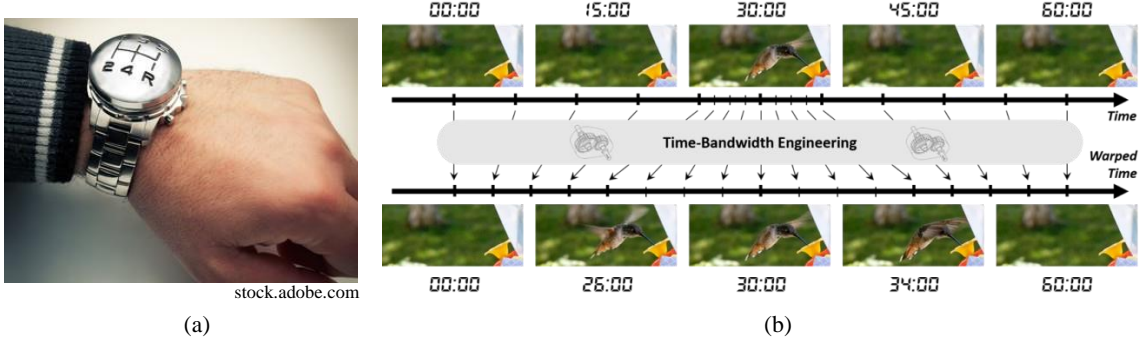

Figure 1: (a) Illustration of the "time gearbox" concept. (b) A time series data experiences a non-uniform stretch transformation such that information rich portions are dilated more than sparse regions[10]. The sampling rate is engineered by tailoring the group velocity dispersion profile of the channel medium in a context-aware fashion. When combined with a fixed-rate ADC, we effectively achieve non-uniform sampling of the input optical spectrum. Here the concept is visualized using a video segment but in practice the transformation is a general warped spectrotemporal or spectrospatial operation; the former is for capture of temporal waveform and the latter for imaging.

* jalali@ucla.edu; photonics.ucla.edu

For extended measurements of ultrafast optical signals, there is a need to tailor the group velocity dispersion of the dispersive fiber in a context-aware manner, such that the total time-bandwidth product of the system can be minimized[11]; recently, it has been shown that one can engineer the time-bandwidth product of data residing on an optical carrier, effectively achieving "photonic hardware acceleration"[10,11]: it is not only possible to slow down the data but also to compress data all-optically[10–13]. This is achieved using warped time stretch implemented with nonlinear group delay dispersion. The device performs variable-rate Fourier domain sampling where the sampling rate self-adapts to the local information content of the signal[10,13].

The function of warped time-stretching can be likened to mechanical gearboxes in vehicles with manual transmission. The sampling rate of the acquisition system can be thought of being the equivalent of the number of teeth per second delivering torque at a certain point on a particular gear, which in turn is related to the rotation speed and the size of the gear used. Just as how a speed mismatch between the car and the gear results in lower power transfer efficiency, and is solved by switching gears, we view our warped Fourier domain sampling approach as an analog gearbox. The warped sampling is achieved by non-uniform mapping of the spectrum into time, such that information rich portions of the spectrum (high local entropy) are stretched more and receive larger number of samples by the ADC than do the sparse regions (low local entropy)[10]. This is done while minimizing the total temporal length (total number of samples), hence compressing the amount of digital data produced.

While this operation is lossless, signal reconstruction (warping) is subject to the signal to noise ratio and renders this a lossy compression technique[12]. The selection of gear ratio to match the acceleration profile of the vehicle is thus equivalent to the selection of dispersion profiles based on local sparsity of the optical spectrum, a quantity that has a mechanical equivalent in the form of "group delay acceleration".

## 2. MATHEMATICAL FORMULATION

To complete the analogy with mechanical gearboxes, we introduce a modal decomposition of the generalized group delay profile. While in principle any basis is valid, we assume a Taylor expansion for the group delay profile $\tau(\omega)$:

$$\tau(\omega) = \sum_{m=1}^{+\infty} \tau_m(\omega) = \sum_{m=1}^{+\infty} \frac{\tau^{(m)}}{(m-1)!}(\omega - \omega_c)^{m-1} \quad (1)$$

where $\tau_m$ is the $m$-th group delay mode, and $\tau^{(m)} \triangleq \frac{d^{m-1}}{d\omega^{m-1}}\tau(\omega)\Big|_{\omega=\omega_c}$ is the $(m-1)$-th order derivative of the group delay evaluated at the centre frequency $\omega_c$, and has units of $\left[\frac{ps}{(Hz)^{m-1}}\right]$. For example, in a fiber of length $L$ with constant group delay dispersion, $\tau^{(1)} = \beta_1 L$ is the overall temporal delay, and the constant $\tau^{(2)} = \frac{d\tau(\omega)}{d\omega}\Big|_{\omega=\omega_c}$ is the traditional definition of the group delay dispersion $\beta_2 L$. In general, each modal component of the group delay inputs a different spectrotemporal characteristic to the optical signal intensity, which collectively affects the output optical field $E_o(t)$ by:

$$E_o(t) = \int_{-\infty}^{+\infty} \tilde{E}_i(\omega - \omega_c) \exp\left(j \int \tau(\omega) d\omega - j(\omega - \omega_c)t\right) \frac{d\omega}{2\pi} \quad (2)$$

where $\tilde{E}_i(\omega - \omega_c)$ is the input optical spectrum and $H(\omega) = e^{i\int \tau(\omega)d\omega}$ describes the operation of the dispersive element[10]. This dispersive element functions as a spectrotemporal gearbox, matching the time-bandwidth product of the input signal to that of the digitizer and subsequent electronics. With highly dispersive elements, the optical spectrum is mapped the into the temporal domain in a process known as time-stretch dispersive Fourier transformation, which mathematically is described by the stationary phase approximation[7,14–16]. Under this approximation, each time point $t$ has a one-to-one correspondence to an optical frequency $\omega_s$ in the optical spectral envelope:

$$\omega_s = \omega_c + \tau^{-1}(t) = \omega_c + \sum_{m=1}^{+\infty} \frac{\Omega^{(m)}}{(m-1)!}\left(t - \tau^{(1)}\right)^{m-1} \quad (3)$$

where $\tau^{-1}(t)$ is the inverse function of the group delay, and $\Omega^{(m)}$ (Taylor series coefficients for the inverse group delay[17]) are the "gears" in the spectrotemporal gearbox:

$$\begin{cases} \Omega^{(1)} = \dfrac{1}{\tau^{(2)}} \\ \Omega^{(2)} = -\dfrac{\tau^{(3)}}{(\tau^{(2)})^3} \\ \Omega^{(3)} = \dfrac{3(\tau^{(3)})^2 - \tau^{(2)}\tau^{(4)}}{6(\tau^{(2)})^5} \\ \ldots \end{cases} \quad (4)$$

The output optical intensity of the signal can then be written as:

$$|E_o(t)|^2 \approx \frac{\left|\tilde{E}_i\left(\omega_c + \tau^{-1}(t)\right)\right|^2}{2\pi} \sum_{m=1}^{+\infty} \frac{\Omega^{(m)}}{(m-2)!}\left(t - \tau^{(1)}\right)^{m-2} \quad (5)$$

With a finite set of group delay modes, we are thus able to design and generate any tailored reshaping that we desire; since the acquisition system "shifts gear" (redistributes the local entropy) automatically from frequency to frequency, **the spectrotemporal gearbox enabled by warped stretch is the optical equivalent of an electrical gearbox, which changes the gear ratios to maximize power transfer efficiency at a given speed.**

## 3. APPLICATIONS

Applied to imaging, time stretch offers real-time image acquisition at millions of frames per second and subnanosecond shutter speed but the system generates massive amount of digital data resulting in a big data bottleneck. The concept of warped stretch has been used to demonstrate all-optical image compression[18], as shown in Figure 2(a-e). Instead of using a variable sampling rate, this was achieved by uniform spectral illumination of the target followed by warped spectro-temporal stretching before the image was digitized by an ADC. The warping was implemented with a chirped fiber Bragg grating boasting nonlinear group delay dispersion. The non-uniform sampling emulates the non-uniform sensor distribution in the retina of the eye, which provides higher resolution in the central vision and coarse but acceptable resolution in the peripheral vision. In contrast, the analog gearbox achieves this by warping the image followed by uniform sampling as opposed to direct non-uniform sampling performed by the eye. Hence, it can operate in real time and without feedback.

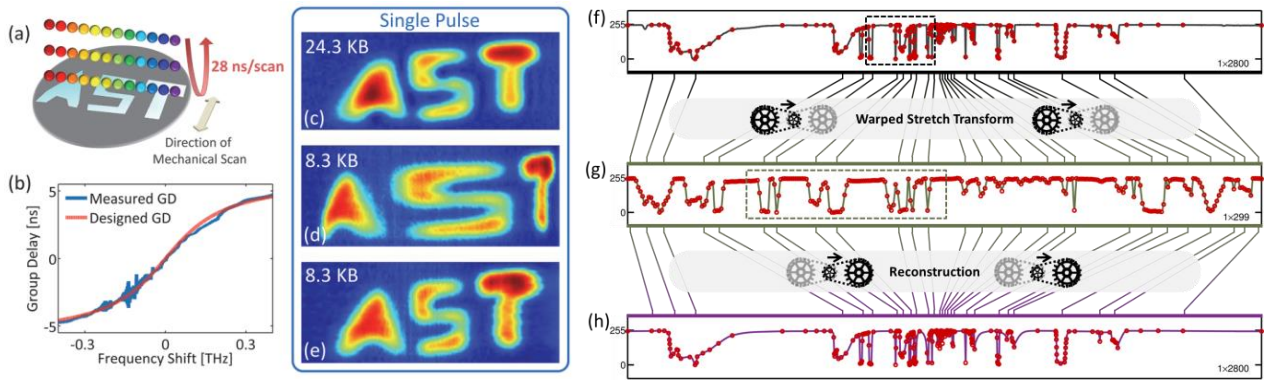

Figure 2: Proof-of-concept experimental results[18]. (a) The test sample reflected one-dimensional rainbow illumination pulses at a scan rate of 36 MHz. (b) The warped stretch transform leading to nonlinear spectrum-to-time mapping is performed by a custom chirped fiber Bragg grating. (c) Image dispersed with a linear group delay profile. (d) Image dispersed by custom chirped fiber Bragg grating. (e) Unwarped image reconstructed from subplot d. (f) Linescan of digital image that is non-uniformly stretched (g) and can be downsampled at a higher rate[19]. (h) Reconstruction of the downsampled stretched signal in subplot g.

It is also possible to implement the warped stretch operations in the numerical domain for applications in digital image processing. The gearbox operation is illustrated in Figure 2(g-i) which shows what happens to a line scan of a digital image[19]. In direct emulation of the analog optical implementation[18] but implemented here in digital domain, the information rich regions of the image are stretched more such that after uniform sampling, one achieves non-uniform sampling with higher sampling rate in the information rich portions. The digital compression method was able to achieve more than 6 dB improvement in the peak signal-to-noise-ratio (PSNR) for 8-bit black-and-white images, when compared to linearly-downsampled images at 4X compression.

## 4. CONCLUSIONS

The analog gearbox is a powerful methodology providing a solution to real-time analog-to-digital conversion of fast optical waveforms while minimizing the amount of data produced by these high throughput systems. outlines a pathway to real-time, single-shot optical computing and signal processing. Much like transmission system in vehicles, we envision the use of dispersive primitives to construct the desired group delay profile by its modal components to play an essential role in spectrotemporal reshaping and encoding of pulse trains in fiber-optic telecommunication systems. The use of tailored dispersive elements in this manner outlines a pathway to real-time, single-shot optical computing and signal processing.

## 5. ACKNOWLEDGMENTS

This work was supported by the Office of Naval Research (ONR) MURI Program on Optical Computing under Award Number N00014-14-1-0505.